\documentclass{aa}  

\usepackage{graphicx}
\usepackage{lscape}
\usepackage{amsmath}
\usepackage{todonotes}
\begin{document}

   \title{BGLS: A Bayesian formalism for the generalised Lomb-Scargle periodogram\thanks{https://www.astro.up.pt/exoearths/tools.html}}

   \subtitle{}

   \author{A. Mortier\inst{1,2},
           J.P. Faria\inst{2,3,4},
           C.M. Correia\inst{2,3},
           A. Santerne\inst{2,3},
        \and
           N.C. Santos\inst{2,3,4}
          }

   \institute{SUPA, School of Physics and Astronomy, University of St Andrews, St Andrews KY16 9SS, UK\\
              \email{am352@st-andrews.ac.uk}
        \and
        Centro de Astrof\'{\i}sica, Universidade do Porto, Rua das Estrelas, 4150-762 Porto, Portugal
        \and
        Instituto de Astrof\'{\i}sica e Ci\^encias do Espa\c co, Universidade do Porto, CAUP, Rua das Estrelas, 4150-762 Porto, Portugal
        \and
        Departamento de F\'{\i}sica e Astronomia, Faculdade de Ci\^encias, Universidade do Porto, Portugal
              }

   \date{Received ...; Accepted ...}

 
  \abstract
   {Frequency analyses are very important in astronomy today, not least in the ever-growing field of exoplanets, where short-period signals in stellar radial velocity data are investigated. Periodograms are the main (and powerful) tools for this purpose. However, recovering the correct frequencies and assessing the probability of each frequency is not straightforward.}
   {We provide a formalism that is easy to implement in a code, to describe a Bayesian periodogram that includes weights and a constant offset in the data. The relative probability between peaks can be easily calculated with this formalism. We discuss the differences and agreements between the various periodogram formalisms with simulated examples.}
   {We used the Bayesian probability theory to describe the probability that a full sine function (including weights derived from the errors on the data values and a constant offset) with a specific frequency is present in the data.}
   {From the expression for our Baysian generalised Lomb-Scargle periodogram (BGLS), we can easily recover the expression for the non-Bayesian version. In the simulated examples we show that this new formalism recovers the underlying periods better than previous versions. A Python-based code is available for the community.}
   {}
   \keywords{ Methods: data analysis -- Methods: statistical
               }

   \authorrunning{Mortier, A. et al.}
   \maketitle
%

\section{Introduction}

Analysing unevenly sampled data in search for a periodic signal is very important in astronomy today. In the ever-growing search for exoplanets, for example, scientists search for periodicities in the radial velocity data of a star to detect a signal that can be attributed to an orbiting planet. When analysing the intensity of stellar activity, we also search for periodic signals, for example in the activity indicator $\log(R'_{HK}),$ or in the photometric light curves \citep[e.g.][]{Dum12,Bast14}. In the field of asteroseismology, correctly finding periodic signals in the data is of great importance as well \citep[e.g.][]{Aerts10}.

From a historical point of view, one of the main tools used in the frequency analysis of unevenly spaced time series is the Lomb-Scargle periodogram \citep[LS -][]{Lomb76,Sca82}. This LS periodogram, although useful, suffers from several drawbacks. First, it does not weigh the data points in any way. Secondly, it does not include a constant offset in the data. These two points, however, are very important when handling real observational data where some data points may be more precise than others due to observing conditions, for instance. The zeropoint of the data is typically not known
exactly, either, which is very important in the case of irregularly sampled data. These problems were already accounted for by e.g. \citet{Fer81}, \citet{Cum99}, \citet{Zech09}, resulting in a generalised LS periodogram (GLS) including weights and an offset.

The final drawback of the LS is also still present in the GLS. Both periodograms are expressed in an arbitrary power, which makes it difficult to compare one peak to another. To better assess the relative probability between two peaks, \citet{Bret01} generalised the LS periodogram by using Bayesian probability theory. The resulting Bayesian LS periodogram (BLS) is in many ways similar to the regular LS periodogram, but the probability resulting from the BLS is much more informative than the arbitrary power in the LS. 

In this work, we follow the formalisms of \citet{Bret01} and \citet{Zech09} to extend the GLS one step further by using Bayesian probability theory. This results in a Bayesian generalised Lomb-Scargle periodogram (BGLS). In Sect. \ref{Form} we derive the equations needed for the BGLS, with a special case described in Sect. \ref{Spec}. Section \ref{Ex} gives two examples based on simulated data to show the differences and agreements between the four discussed periodograms (LS, BLS, GLS, BGLS). We conclude in Sect. \ref{Con}.

\section{Formalism}\label{Form}

A periodic signal in time series data can be described by a full sine function, including an offset and the errors on the observations. The model we used for the data is

\begin{equation}\label{eqmod}
d(t_i) = d_i = A\cos(2\pi ft_i - \theta) + B \sin(2\pi ft_i - \theta) + \gamma + \epsilon_i,
\end{equation}

\noindent where $d_i$ is the data point taken at time $t_i$, $A$ and $B$ are the cosine and sine amplitudes, $f$ is the signal frequency, $\theta$ is an arbitrary phase offset that we defined below, $\gamma$ is the data offset, and $\epsilon_i$ is the noise at time $t_i$. This noise is per time $t_i$ Gaussian-distributed around $0$ with a standard deviation of $\sigma_i$, which is the estimated uncertainty on the data at time $t_i$ $(\epsilon_i\sim N(0,\sigma_i))$.

We are interested in the posterior probability of the frequency given the data $D$ and our prior knowledge $I$, $P(f|D,I)$. By marginalisation, we can write this (posterior) probability function as

\begin{equation}
P(f|D,I) = \int\int\int P(fAB\gamma|D,I)\,dA\,dB\,d\gamma,
\end{equation}

\noindent where the frequency probability is calculated from the joint posterior probability of the three parameters $A$, $B$, and $\gamma$. By applying Bayes theorem \citep{Bay1763}, we can write that

\begin{equation}
P(fAB\gamma|D,I) = \frac{P(D|fAB\gamma,I)P(fAB\gamma|I)}{P(D|I)}.
\end{equation}

The evidence, $P(D|I)$, is a constant, and since we are only interested in the relative probability between frequencies, we can ignore this normalising factor. Furthermore, for the priors, $P(fAB\gamma|I)$, we assume that all the parameters $f,A,B$, and $\gamma$ are independent, and we take the prior probability for each parameter as uniform. This assumption leads to an equal treatment of all possible signals in the data. With these assumptions, we can use the product rule for joint probabilities: 

\begin{equation}
P(fAB\gamma|I) = P(f|I)P(A|I)P(B|I)P(\gamma|I) = ct,
\end{equation}

\noindent with $ct$ a constant that we can again ignore. This all leads to the fact that our posterior probability function is proportional to the integrated likelihood:

\begin{equation}\label{EqIntLH}
P(f|D,I) \propto \int\int\int P(D|fAB\gamma,I)\,dA\,dB\,d\gamma. 
\end{equation}

To describe this likelihood analytically, we use the fact that the probability of the data is the same as the probability of the noise. This noise is normally distributed around $0$, with the standard deviation $\sigma_i$ for each time $t_i$. We derive

\begin{eqnarray}
P(D|fAB\gamma,I) & = & \prod_{i=1}^N \frac{1}{\sqrt{2\pi}\sigma_i} \exp\left(-\frac{\epsilon_i^2}{2\sigma_i^2} \right)\\
 & = & \left( \prod_{i=1}^N \frac{1}{\sqrt{2\pi}\sigma_i} \right) \exp\left[-\frac{1}{2}\sum_{i=1}^N\left(\frac{\epsilon_i}{\sigma_i} \right)^2\right].
\end{eqnarray}

\noindent The first factor in this expression can again be ignored and placed into the normalising factor. We use the standard deviations $\sigma_i$ to assign the weights of the datapoints. The weight $w_i$ can be written as

\begin{equation}
w_i  =  \frac{1}{\sigma_i^2}
.\end{equation}

The noise $\epsilon_i$ can be written in terms of the observables by using Eq. \ref{eqmod}. To simplify the expressions, we introduce some definitions. We try to use the same expressions as in \citet{Zech09} so that it is easy to compare between the two formalisms.

\begin{eqnarray}
W & = & \sum_{i=1}^N w_i \\
Y & = & \sum_{i=1}^N w_i d_i\\
\widehat{YY} & = & \sum_{i=1}^N w_i d_i^2\\
\widehat{YC} & = & \sum_{i=1}^N w_i d_i \cos(2\pi ft_i - \theta) \\
\widehat{YS} & = & \sum_{i=1}^N w_i d_i \sin(2\pi ft_i - \theta) \\
C & = & \sum_{i=1}^N w_i \cos(2\pi ft_i - \theta) \\
S & = & \sum_{i=1}^N w_i \sin(2\pi ft_i - \theta) \\
\widehat{CC} & = & \sum_{i=1}^N w_i \cos^2(2\pi ft_i - \theta) \\
\widehat{SS} & = & \sum_{i=1}^N w_i \sin^2(2\pi ft_i - \theta)
.\end{eqnarray}

Furthermore, we define $\theta$ such that the cosine and sine functions are orthogonal (for the proof, see Appendix \ref{theta}):

\begin{equation}\label{EqTheta}
\theta = \frac{1}{2}\tan^{-1}\left[\frac{\sum w_i\sin(4\pi f t_i)}{\sum w_i\cos(4\pi f t_i)} \right]
.\end{equation}

Using all these definitions, we find that

\begin{equation}
\begin{split}
\sum_{i=1}^N\left(\frac{\epsilon_i}{\sigma_i} \right)^2 = & \widehat{YY} - 2A\widehat{YC} - 2B\widehat{YS} - 2\gamma Y + A^2\widehat{CC} \\ 
& + B^2\widehat{SS} + \gamma^2W + 2A\gamma C + 2B\gamma S
\end{split}
.\end{equation}

Using this expression, we can now split the integrals from Eq. \ref{EqIntLH} into three separate integrals. First, we consider the integrals in $A$ and $B$. These can be written as

\begin{eqnarray}
\int_{-\infty}^{+\infty} \exp\left[- \frac{\widehat{CC}A^2-2\widehat{YC}A+2\gamma C A}{2} \right] dA \\
\int_{-\infty}^{+\infty} \exp\left[- \frac{\widehat{SS}B^2-2\widehat{YS}B+2\gamma S B}{2} \right] dB .
\end{eqnarray}

From now on, we assume that $\widehat{CC}$ and $\widehat{SS}$ are strictly positive. Since $\cos^2x\geq0$ and $\sin^2x\geq0$, it follows that these expressions are positive. We explore the special cases where one of them is zero in Sect. \ref{Spec}. If they are strictly positive, we can easily solve the integrals in $A$ and $B$ using

\begin{equation}\label{EqInt}
\int_{-\infty}^{+\infty} e^{-ax^2} e^{-2bx} dx = \sqrt{\frac{\pi}{a}} \exp\left(\frac{b^2}{a}\right).
\end{equation}

Finally, only the integral in $\gamma$ remains, and our posterior probability function is proportional to

\begin{equation}\label{EqExpKLM}
P(f|D,I) \propto \frac{1}{\sqrt{\widehat{CC}\widehat{SS}}} \int \exp(K\gamma^2 + L\gamma + M) d\gamma,
\end{equation}

\noindent where we have introduced the following functions of the frequency $f$:

\begin{eqnarray}
K & = & \frac{C^2\widehat{SS} + S^2\widehat{CC} - W\widehat{CC}\widehat{SS}}{2\widehat{CC}\widehat{SS}} \\
L & = & \frac{Y\widehat{CC}\widehat{SS} - C\widehat{YC}\widehat{SS} - S\widehat{YS}\widehat{CC}}{\widehat{CC}\widehat{SS}} \\
M & = & \frac{\widehat{YC}^2\widehat{SS} + \widehat{YS}^2\widehat{CC}}{2\widehat{CC}\widehat{SS}} 
.\end{eqnarray}

To solve the integral in Eq. \ref{EqExpKLM}, we can again use Eq. \ref{EqInt} since $K<0$. For data sets with $N\geq3$ (a frequency analysis is only useful for datasets with at least three datapoints) and unevenly spaced datapoints, this is always the case. We then finally obtain that

\begin{equation}\label{EqBGLS}
P(f|D,I) \propto \frac{1}{\sqrt{|K|\widehat{CC}\widehat{SS}}} \exp\left( M - \frac{L^2}{4K}\right)
.\end{equation}

As can be seen from Eq. 20 in \citet{Zech09}, the power calculated in the GLS is proportional to $M$. The probability calculated by the BGLS is related to the GLS as it gives an exponential where part of the exponent is exactly the power from the GLS. The BGLS is thus very similar to the GLS, just as the BLS is similar to the LS \citep[see discussion in ][]{Bret01}.

\section {Special case}\label{Spec}

In the previous section, we have assumed that $\widehat{CC}$ and $\widehat{SS}$ are strictly positive. Now, we consider the
case when one of them is zero\footnote{They cannot be zero at the same time.}. This can be the case if, for a specific frequency $f$, all datapoints $t_i$ can be expressed as

\begin{equation}
2ft_i = 2ft_1 + k, k\in \mathbb{Z}.
\end{equation}

Basically, this means that the data are equally spaced, while gaps without observations are allowed. From Eq. \ref{EqTheta} we can then derive that

\begin{equation}
\theta = 2\pi ft_1 + k\frac{\pi}{2}, k\in \mathbb{Z}.
\end{equation}

By placing these two expressions together, we obtain that

\begin{equation}\label{EqSpec}
2\pi ft_i - \theta = k\frac{\pi}{2}, k\in \mathbb{Z}.
\end{equation}

Now, $\widehat{CC}$ and $\widehat{SS}$ will be zero if $k$ is odd or even, respectively. If $k$ is even, we see that

\begin{equation}
\widehat{SS} = S = \widehat{YS} = 0.
\end{equation}

With these simplified expressions, we find that the integral in $B$ is proportional to a constant. The integral in $A$ can still be solved, and we can follow the same reasoning as in the previous section. Accordingly, the BGLS can be expressed as

\begin{equation}
P(f|D,I) \propto \frac{1}{\sqrt{|K_C|\widehat{CC}}} \exp\left( M_C - \frac{L_C^2}{4K_C}\right)
,\end{equation}

\noindent where we have used the following definitions:

\begin{eqnarray}
K_C & = & \frac{C^2 - W\widehat{CC}}{2\widehat{CC}} \\
L_C & = & \frac{Y\widehat{CC} - C\widehat{YC}}{\widehat{CC}} \\
M_C & = & \frac{\widehat{YC}^2}{2\widehat{CC}} .
\end{eqnarray}

For the case where $k$ in Equation \ref{EqSpec} is odd, we can reason the same way, but now all the terms with cosines are zero and the terms with sines are left (introducing the three functions $K_S,L_S,$ and $M_S$).

These cases may occur for equally spaced data sets (even if they have gaps). With real ground-based observations, it is very unlikely that these cases would ever occur. However, this is more likely
for space-based surveys, where the observations are made at fixed times following the spacecraft clock. Therefore the frequencies for which these special cases apply need
to be checked always.

\section{Simulated examples}\label{Ex}

In this section we provide two examples of simulated data sets that expose the differences between the LS, the GLS and their Bayesian versions. 

The main difference between the non-Bayesian and the Bayesian periodograms lies in the peak comparison, as already mentioned in the introduction. Non-Bayesian periodograms use an arbitrary power, which makes it difficult to assess the importance of specific periods over other periods in the data. Bayesian periodograms, on the other hand, express the probability that a signal with a specific period is present in the data. The relative probability between two periods can then
be easily assessed. In the examples below, we show that the Baysian periodograms typically have one very clear peak (probability-wise), in contrast to their non-Bayesian versions.

Differences between the non-generalised and the generalised periodograms always have to do with the fact that the (B)GLS includes weights and a possible offset in the data. Below, we give an example of the effect of each of these additions using simulated data\footnote{We chose to label the ordinates as \textit{RV} and they are expressed in arbitrary units.}. For every example we provide the LS and GLS expressed in the arbitrary power and the BLS and BGLS expressed in a normalised probability (so that the highest peak represents $100\%$ probability). We show the Bayesian periodograms on a linear scale and on a logscale, where it is clearer that they are indeed linked to their non-Bayesian versions. Note that the plots on logscale show a horizontal offset between the BLS and BGLS. This is due to the normalisation of the highest peak to $100\%$.

   \begin{figure}
   \centering
   \includegraphics[width=\linewidth]{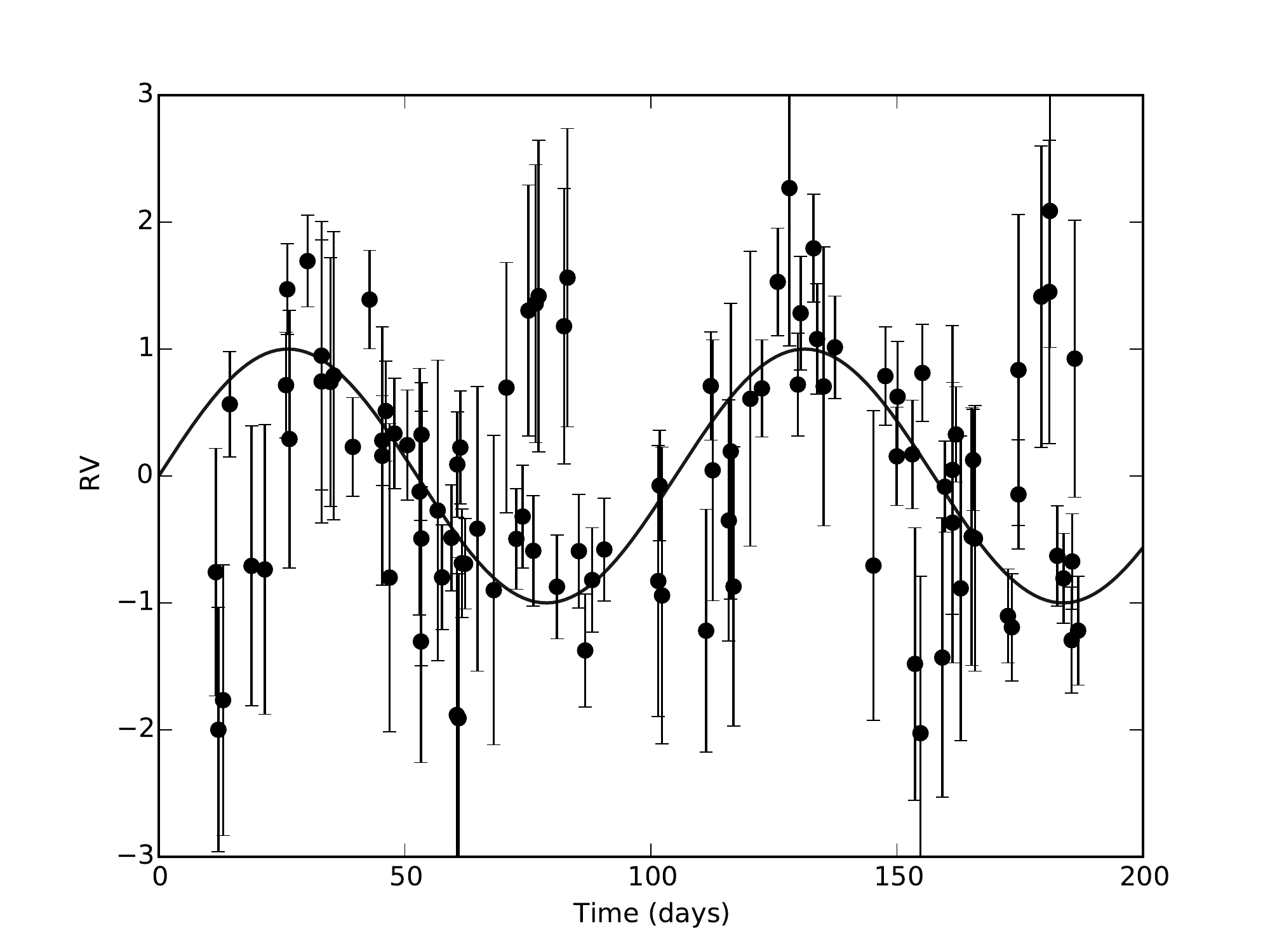}
      \caption{Simulated data (black points) to show the effect of weighting the datapoints using the errorbars. The real underlying signal with a period of 105 days and a semi-amplitude of 1 is represented by the black curve.}
         \label{fig:ex1_obs}
   \end{figure}

   \begin{figure}
   \centering
   \includegraphics[width=\linewidth]{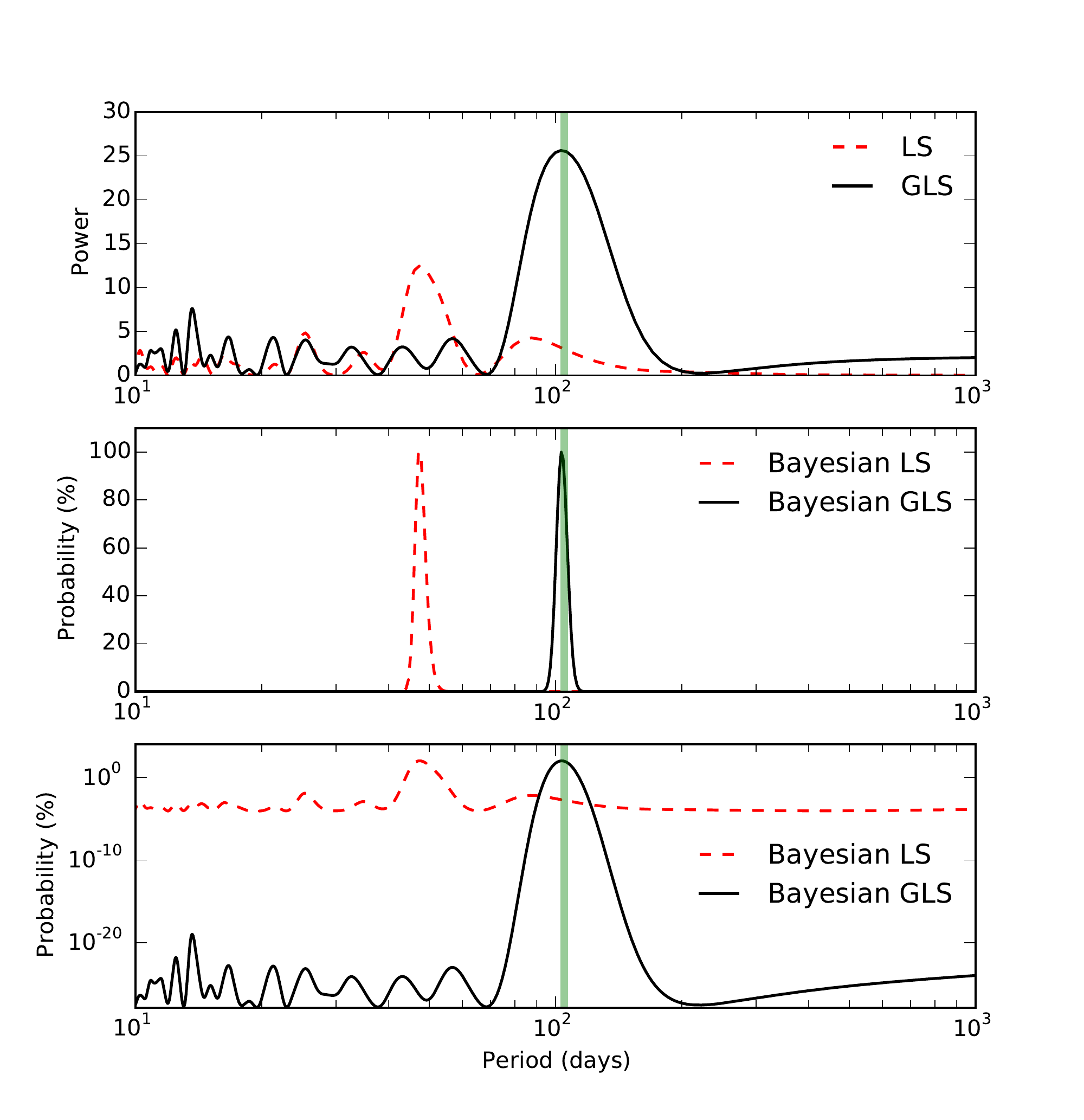}
      \caption{Periodograms of the simulated data from Fig. \ref{fig:ex1_obs}. The top panel shows the non-Bayesian versions, the middle and bottom panels show the Bayesian versions on a linear and logscale.}
         \label{fig:ex1_per}
   \end{figure}

To consider the weighting of the data points, we simulated $100$ unevenly spaced data points over a range of $180$ days. The real underlying signal in these data has a period of $105$ days and a semi-amplitude of $1$. Half of the points have a mean error
bar of $0.4,$  the other half have a mean error bar of $1.1$. Furthermore, by this addition of random white noise, the less precise points also deviate more from the underlying periodic signal. In real observations, this can easily occur because the precision of measurements strongly depends on the observing conditions (e.g. seeing, exposure time, instrument setup), and periodic signals hidden in the data often have semi-amplitudes of the order of the error bars.

The simulated data and the real underlying signal are shown in Fig. \ref{fig:ex1_obs}. The resulting periodograms are shown in Fig. \ref{fig:ex1_per}. While the 105-day period signal is clearly detected by the (B)GLS, the (B)LS does not single out
that period as the most significant. Instead, it settles on a
period of around 50 days. In both the non-Bayesian and Bayesian periodograms, this is clear. Furthermore, the second peak in the (B)LS lies not exactly at 105 days either, but more towards a slightly shorter period.

   \begin{figure}
   \centering
   \includegraphics[width=\linewidth]{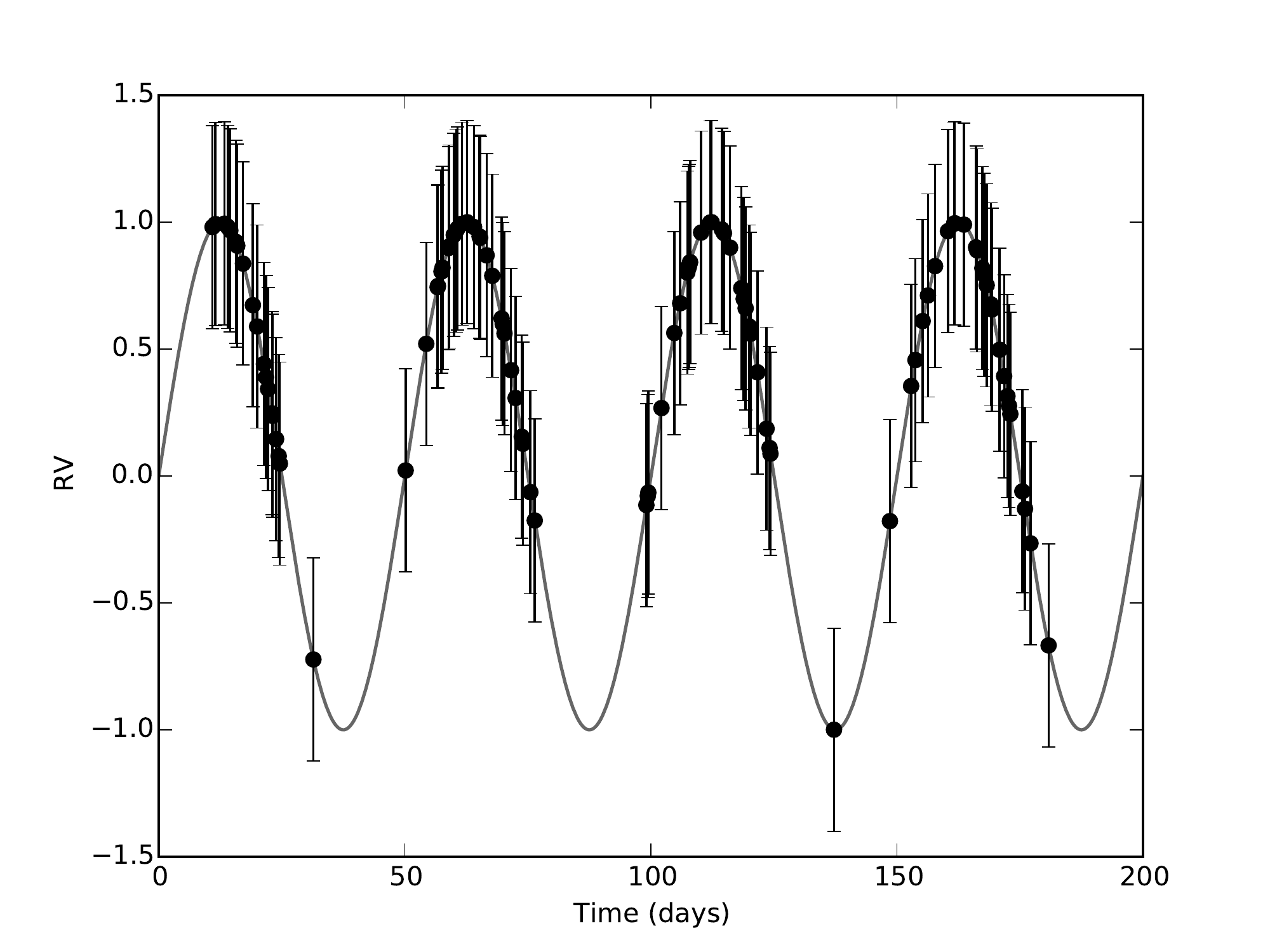}
      \caption{Simulated data (black points) to show the effect of including an offset in the analysis. The real underlying signal with a period of 50 days and a semi-amplitude of 1 is represented by the black curve.}
         \label{fig:ex2_obs}
   \end{figure}

   \begin{figure}
   \centering
   \includegraphics[width=\linewidth]{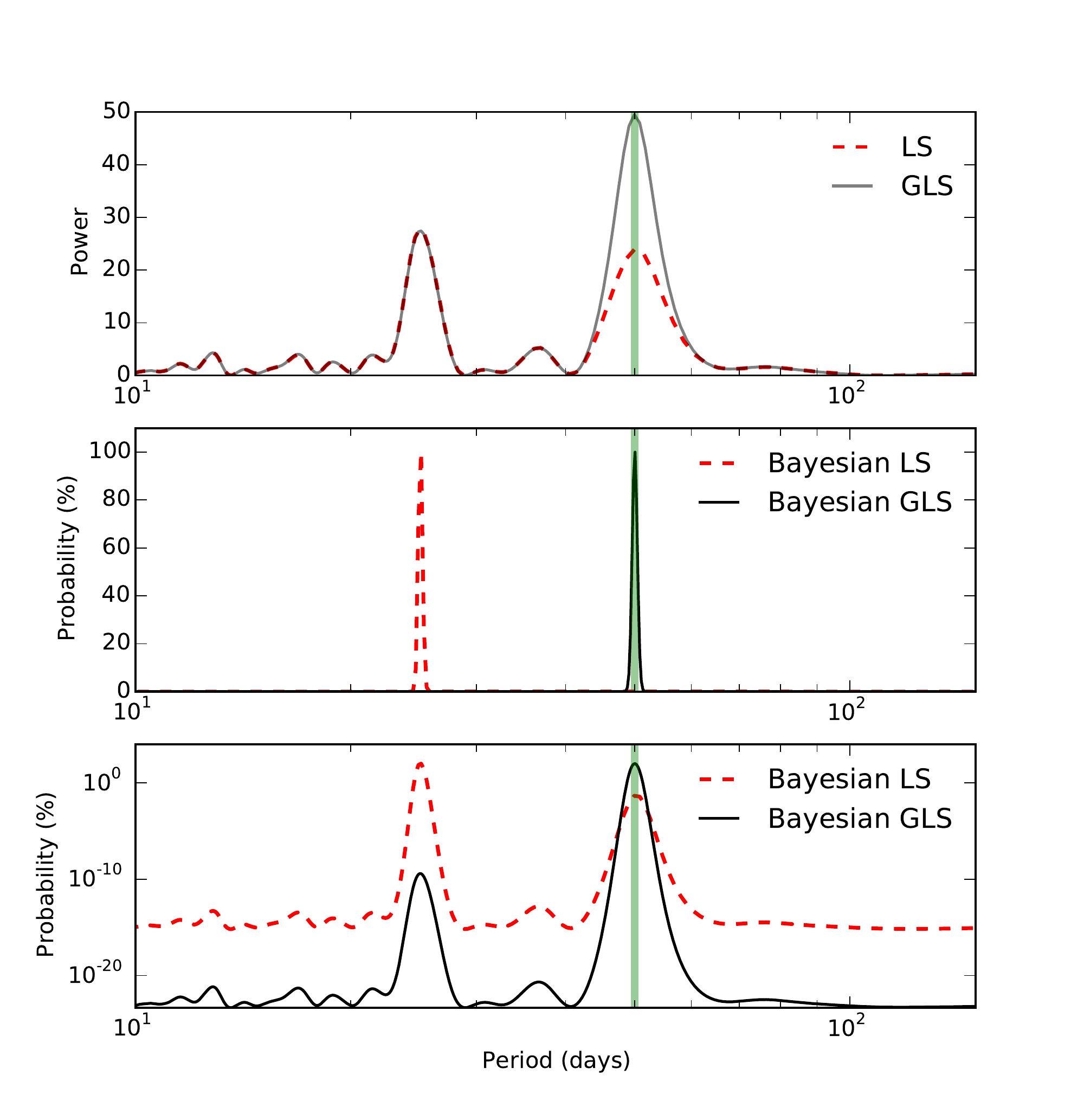}
      \caption{Periodograms of the simulated data from Fig. \ref{fig:ex2_obs}. The top panel shows the non-Bayesian versions, the middle and bottom panels show the Bayesian versions on a linear and logscale.}
         \label{fig:ex2_per}
   \end{figure}

Including a constant offset in the data analysis allows for the mean of the data to be different from the mean of the fitted sine function. These two means can be very different depending on the sampling of the data. For real observations, especially if they are ground-based, the sampling is constrained by the observation opportunities (night versus day, weather conditions, seasonal visibility, etc.). To illustrate the effect of this periodic sampling and thus the inclusion of an offset, we simulated $100$ unevenly spaced data points over a range of $170$ days. The real underlying signal in these data has a period of $50$ days, a semi-amplitude of $1,$ and a mean of zero. The data were chosen so that most points lie above $-0.3$ thus creating large gaps in the data of almost no observations.

The simulated data are shown in Fig. \ref{fig:ex2_obs} together with the underlying sine function. The data have a mean of $0.5484$, substantially different from the mean of the underlying sine function, which is zero. The resulting periodograms are shown in Fig. \ref{fig:ex2_per}. It is immediately clear that only the GLS and BGLS are able to correctly identify the true period. It is also interesting to note that the GLS shows two peaks of about the same height at $50$ and $25$ days. By using the GLS
alone, it would not be possible
to know unambiguously which period is the real one in the data. In the BGLS, on the other hand, it is very clear that the longer period (which is the true one) is about $10^10$ times more probable than the other period.

If the data are randomly sampled and include an offset, the mean of the data will be close enough to the value of this offset so that it can easily be subtracted. In this case, all four periodograms we discussed will identify the correct period. More details can be found in Appendix \ref{ExExtra}.

\section{Conclusion}\label{Con}

We formulated an expression for calculating the Bayesian generalised Lomb-Scargle periodogram (given in Eq. \ref{EqBGLS}). Part of the exponent in that expression is the power for the GLS as given by \citet{Zech09}, clearly linking the two expressions. A Python-based code is available for the community at \url{https://www.astro.up.pt/exoearths/tools.html}.

This expression is valid for all unevenly sampled datasets for which Eq. \ref{EqSpec} does not hold for all even or all odd values of $k$. If these equations would hold anyway, we provided the corresponding (simplified) expressions in Sect. \ref{Spec}. 

We repeat that this formalism was reached by making some assumptions. First, we assumed that the parameters $f$, $A$, $B$, and $\gamma$ are independent. Second, we assumed uniform priors for these parameters. The same assumptions were made in \citet{Bret01}. If different assumptions were used, the formalisms in either works are no longer valid.

We simulated two datasets to show the differences between the LS, BLS, GLS, and BGLS. We showed that the generalised periodograms work best for recovering the correct frequencies. Furthermore, we showed that for two peaks with similar heights in the GLS, the BGLS can clarify which peak is the more probable and by how much. We therefore conclude that the BGLS is the most powerful way to explore periodicities in unevenly spaced datasets.

As a final note, we caution that the four periodograms we described and used all assume a single-frequency signal in the data. If multiple frequencies are present in the data (with similar or different amplitudes), all of these formalisms can fail in detecting the correct frequencies. There are some codes in the literature to make multi-frequency periodograms \citep[e.g.][]{Balu13}. However, these also rely on single-frequency periodograms for an initial assessment of the possible frequencies since correctly calculating a multi-frequency periodogram needs too much computing time \citep{Balu13}. Furthermore, we note that a periodogram gives no information about the physical nature of the signal it detects, and one should always be careful with using the periodogram output.

\begin{acknowledgements}

The research leading to these results received funding from the European Union Seventh Framework Programme (FP7/2007-2013) under grant agreement number 313014 (ETAEARTH).
We also acknowledge the support from Funda\c{c}\~ao para a Ci\^encia e a Tecnologia (FCT, Portugal) through FEDER funds in program COMPETE, as well as through national funds, in the form of grants reference RECI/FIS-AST/0176/2012 (FCOMP-01-0124-FEDER-027493) and RECI/FIS-AST/0163/2012 (FCOMP-01-0124-FEDER-027492). We also acknowledge the support from the European Research Council/European Community under the FP7 through Starting Grant agreement number 239953 and the Marie Curie Intra-European Fellowship with reference FP7-PEOPLE-2011-IEF, number 300162 (for CC).
NCS was supported by FCT through the Investigador FCT contract reference IF/00169/2012
and POPH/FSE (EC) by FEDER funding through the program "Programa Operacional de Factores de Competitividade - COMPETE. 
A.S. is supported by the European Union under a Marie Curie Intra-European Fellowship for Career Development with reference FP7-PEOPLE-2013-IEF, number 627202.

\end{acknowledgements}

\bibliographystyle{aa} 
\bibliography{/home/annelies/My_Articles/References.bib}

\appendix

\section{Proof for equation \ref{EqTheta}}\label{theta}

If the cosine and sine functions are orthogonal, it holds that

\begin{equation}\label{EqStart}
\sum_{i=1}^N w_i \sin(2\pi ft_i - \theta) \cos(2\pi ft_i - \theta) = 0.
\end{equation}

We can evaluate the sine and the cosine by using the trigonometric addition formulae:

\begin{eqnarray}
\sin(2\pi ft_i - \theta) = \sin(2\pi ft_i)\cos\theta - \cos(2\pi ft_i)\sin\theta \\
\cos(2\pi ft_i - \theta) = \cos(2\pi ft_i)\cos\theta + \sin(2\pi ft_i)\sin\theta
.\end{eqnarray}

The product of the sine and cosine then becomes

\begin{equation}
\begin{split}
\sin(2\pi ft_i)\cos(2\pi ft_i)\cos^2\theta + \sin^2(2\pi ft_i)\sin\theta\cos\theta \\
- \sin(2\pi ft_i)\cos(2\pi ft_i)\sin^2\theta - \cos^2(2\pi ft_i)\sin\theta\cos\theta .
\end{split}
\end{equation}

Now we can use the double-angle formulae to derive that Eq. \ref{EqStart} can be written as

\begin{equation}
\sum_{i=1}^N w_i \sin(4\pi ft_i)\cos(2\theta) = \sum_{i=1}^N w_i \cos(4\pi ft_i)\sin(2\theta) 
.\end{equation}

From here it easily follows that

\begin{equation}
\tan(2\theta) = \frac{\sum w_i\sin(4\pi f t_i)}{\sum w_i\cos(4\pi f t_i)}
,\end{equation}

and thus that Eq. \ref{EqTheta} holds,

\begin{equation}
\theta = \frac{1}{2}\tan^{-1}\left[\frac{\sum w_i\sin(4\pi f t_i)}{\sum w_i\cos(4\pi f t_i)} \right]
.\end{equation}

\section{Example with offset on randomly sampled data}\label{ExExtra}

To illustrate the effect of an offset if the data are randomly sampled, we include here an additional example. We simulated $100$ unevenly spaced, but randomly sampled data points over a range of $180$ days. The real underlying signal in these data has a period of $50$ days, a semi-amplitude of $1,$ and an offset of $2.5$. To calculate the LS and the BLS, the mean of the data is subtracted first. In this case, the mean is $2.42$, thus very close to the offset value. 

\begin{figure}
   \centering
   \includegraphics[width=\linewidth]{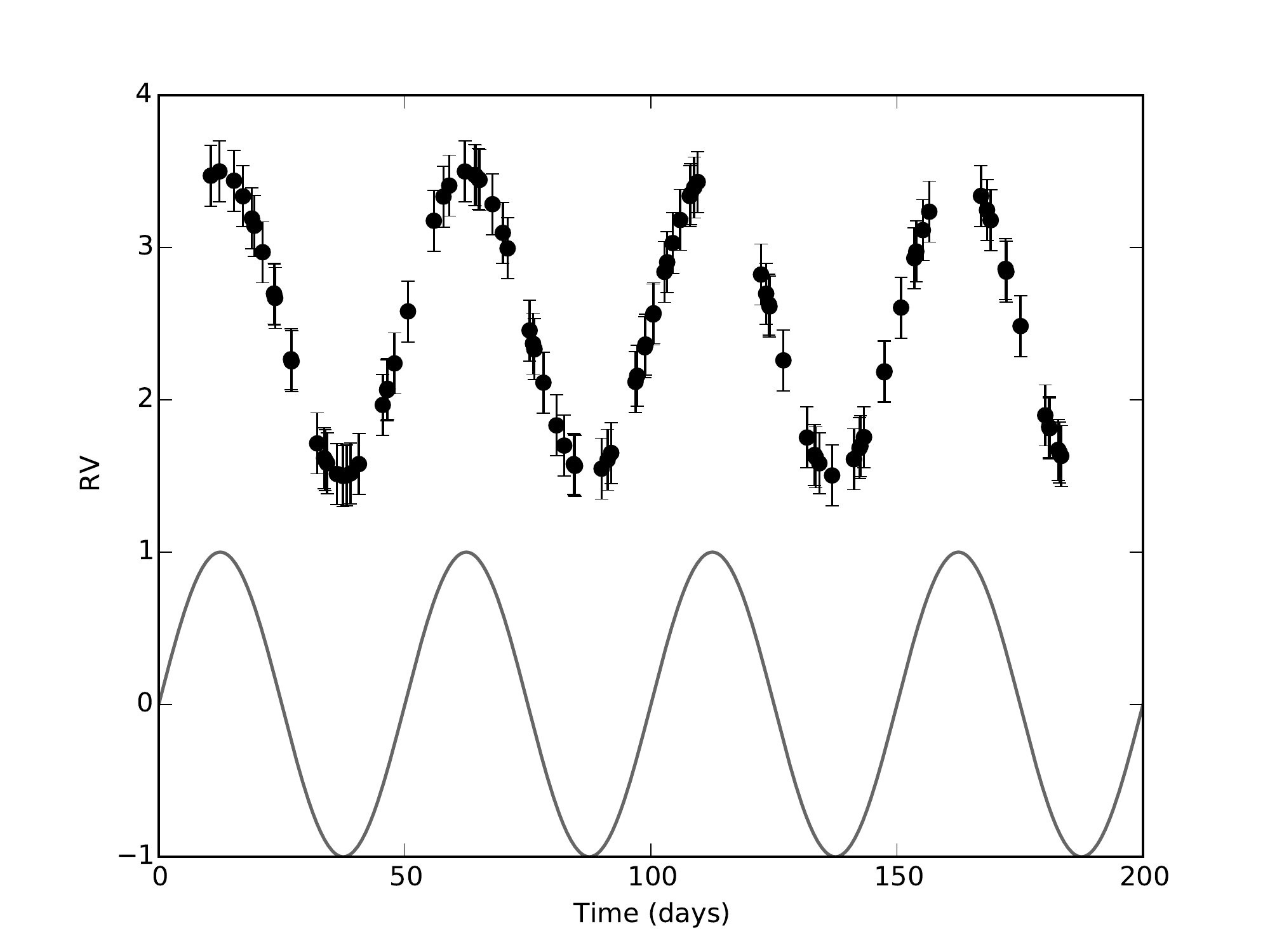}
      \caption{Simulated data (black points) to show the effect of an offset (of 2.5) on randomly sampled data. The underlying signal with a period of 50 days and a semi-amplitude of 1 is represented without the offset (black curve).}
         \label{fig:exap_obs}
   \end{figure}

\begin{figure}
   \centering
   \includegraphics[width=\linewidth]{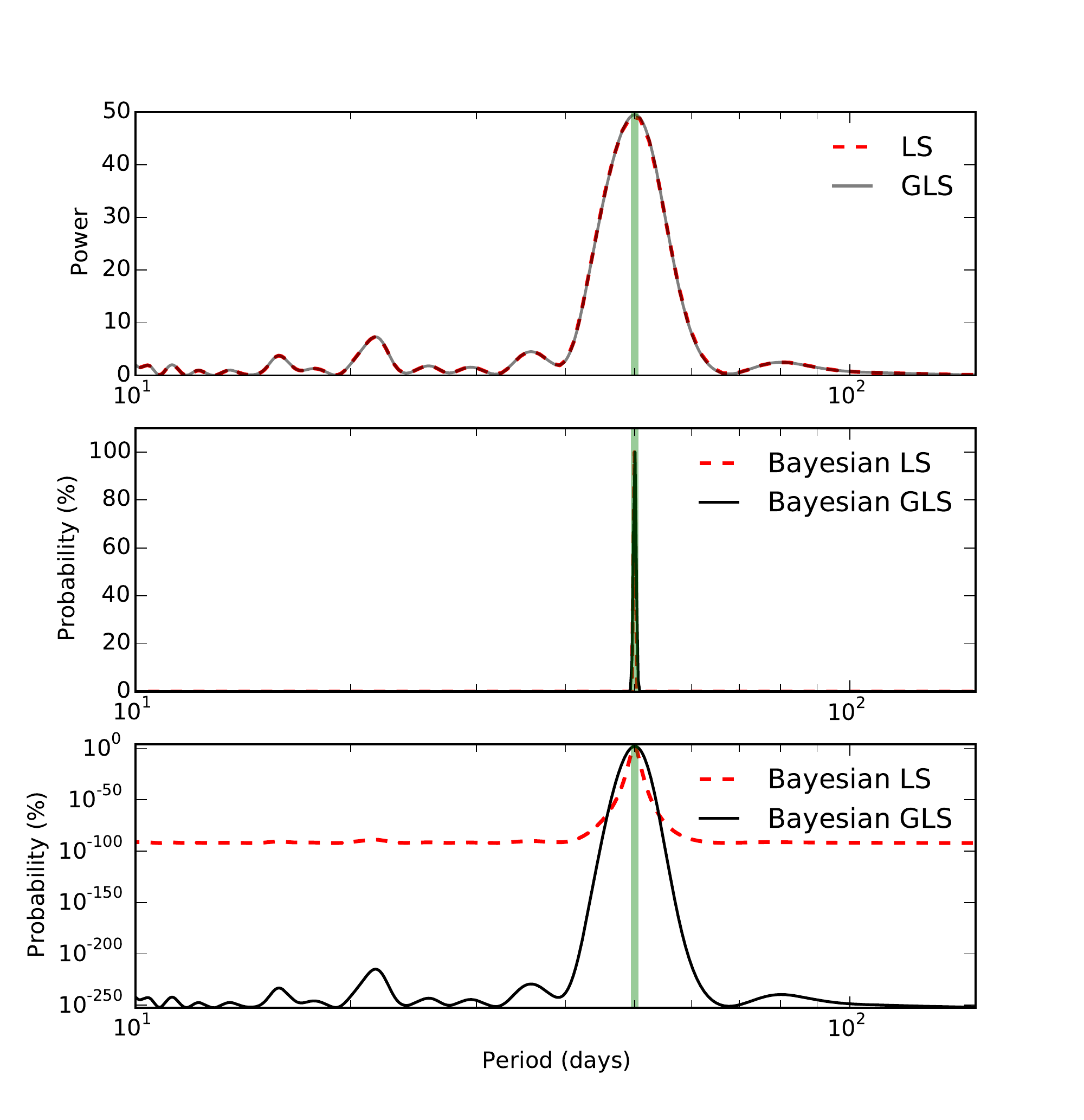}
      \caption{Periodograms of the simulated data from Fig. \ref{fig:exap_obs}. The top panel shows the non-Bayesian versions, the middle and bottom panel show the Bayesian versions on a linear and logscale.}
         \label{fig:exap_per}
   \end{figure}

The simulated data, including the offset, is shown in Fig. \ref{fig:exap_obs} together with the underlying sine function, which includes no offset. The resulting periodograms are shown in Fig. \ref{fig:exap_per}. As expected, the four periodograms all identify the same (and correct) period. From the Bayesian periodograms, it is clear that the identified period is more probable using the BGLS than it is using the BLS (though both are highly significant). This is because the mean of the data is very close to the offset value, but not exactly the offset value.

\end{document}